\documentclass[11pt]{article}
\usepackage{geometry}
\usepackage{here}
\usepackage{graphicx}
\usepackage{amsmath,amssymb}
\usepackage{cite}
\usepackage{bm}
\usepackage[doublespacing]{setspace}
\newcommand{\ra}[1]{\smash{\raise 2.0ex\hbox{#1}}}
\newcommand{\lw}[1]{\smash{\lower 1.5ex\hbox{#1}}}

\newcommand{\mapright}[1]{\smash{\mathop{\hbox to 1cm{\rightarrowfill}}\limits^{#1}}}

\begin{document}
\title{
Glauber--Lachs formula-based analysis of three-pion Bose--Einstein correlation data in $pp$ collisions at 7 TeV from the LHCb Collaboration}

\author{Takuya Mizoguchi$^{1}$\thanks{E-mail: mizoguti@toba-cmt.ac.jp}, Seiji Matsumoto$^{2}$, and Minoru Biyajima$^{3}$\\
{\small $^{1}$National Institute of Technology, Toba College, Toba 517-8501, Japan}\\
{\small $^{2}$Data Science Education Headquarters, Organization for Education and Student Welfare,}\\ {\small Shinshu University, Matsumoto 390-8621, Japan}\\
{\small $^{3}$Center for General Education, Shinshu University, Matsumoto 390-8621, Japan}}

\date{}
\maketitle

\begin{abstract}
In this study, we combine the Glauber--Lachs formula from quantum optics and the two-component picture for pion production to analyze data on two- and three-pion Bose--Einstein correlation in $pp$ collisions at 7 TeV from the LHCb Collaboration. For the pion exchange function $E_{\rm 2B}$, we chose a dipole form and an inverse one-and-a-half pole form. The extensions are computed in the configuration space of four-dimensional Euclidean space ($\xi=\sqrt{|\bm r_1-\bm r_2|^2+(t_1-t_2)^2}$).
\end{abstract}

\section{\label{sec1}Introduction}

\begin{figure}[H]
  \centering
  \includegraphics[width=0.25\columnwidth]{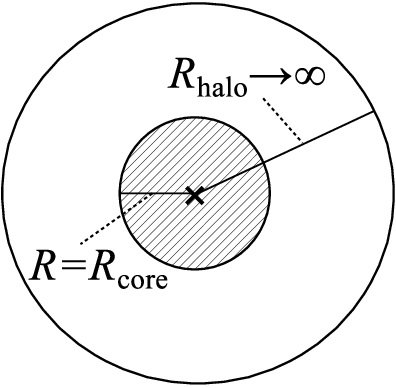}
  \caption{\label{fig1} Schematic of the core/halo model with $E_{\rm 2B}=e^{-RQ}$. Note that $\langle n_{\rm core}\rangle$ contains the chaotic component $A_c$ and the coherent component $|\zeta_c|^2$.}
\end{figure}

Recently, the LHCb Collaboration reported their data on three-pion Bose--Einstein correlation (BEC) in $pp$ collisions at 7 TeV~\cite{LHCb:2025jci}. In this context, explorations of the Hanbury Brown and Twiss effect~\cite{HanburyBrown:1956bqd} in pions and Goldhaber--Goldhaber--Lee--Pais effect~\cite{Goldhaber:1960sf} remain relevant. The LHCb Collaboration used the Glauber--Lachs (GL) formula~\cite{Glauber:1966,Lachs:1965zz,Biyajima:1978cz,Gyulassy:1979yi,Biyajima:1979ak,Biyajima:1982un,Biyajima:1990ku,UA1-MinimumBias:1991afz,Andreev:1991eu,LHCb:2017pnz} with the ratio $f_c=\langle n_{\rm core}\rangle/\langle n_{\rm tot}\rangle$ in the core/halo model shown in Fig.~\ref{fig1}, where $\langle n_{\rm tot}\rangle = \langle n_{\rm core}\rangle+\langle n_{\rm halo}\rangle$. In the original GL formula the average pion multiplicity $\langle n_{\rm tot}\rangle$ is described by a chaotic component $A$ and a coherent component $|\zeta|^2$, such that $\langle n_{\rm tot}\rangle = A+|\zeta|^2$. The identities $p=A/\langle n_{\rm tot}\rangle$ and $(1-p)=|\zeta|^2/\langle n_{\rm tot}\rangle$ relate to the chaotic and coherent components. The LHCb Collaboration analyzed their data with the replacement: $f_cp_c=f_c|\zeta_c|^2/\langle n_{\rm core}\rangle$ corresponding to $(1-p)$ in original GL formula, with $f_c$ and the coherent component in core $|\zeta_c|^2$. Thus, the formula for data on two-pion BEC is expressed as follows:
\begin{eqnarray}
\left.\frac{N^{(2\pi)}}{N^{\rm BG}}\right|_{\rm LHCb} = (1 + \lambda e^{-RQ})\times C(1+\delta Q),
\label{eq1}
\end{eqnarray}
where $\lambda = f_c^2[(1-p_c)^2+2(1-p_c)p_c]$~\cite{LHCb:2025jci}, and $Q$ is four-momentum transfer square\\ $Q=\sqrt{|\bm p_1-\bm p_2|^2-(p_{10}-p_{20})^2}$. $R$ defines the range of the pion production region. In~\cite{LHCb:2025jci}, Eq.~(\ref{eq1}) was applied to data on two-pion BEC from 2017~\cite{LHCb:2017pnz}. 

The data on three-pion BEC is described by the following equation with $3l_2+l_3=\lambda_3$, 
\begin{eqnarray}
\left.\frac{N^{(3\pi)}}{N^{\rm BG}}\right|_{\rm LHCb} = (1 + 3l_2 e^{-RQ} + l_3 e^{-\frac 32RQ})\times C(1+\delta Q)^3,
\label{eq2}
\end{eqnarray}
where $\lambda_3 = 3f_c^2[(1-p_c)^2+2(1-p_c)p_c]+2f_c^3[(1-p_c)^3+3(1-p_c)^2p_c]$, and $Q_{3\pi}^2 = Q_{12}^2+Q_{23}^2+Q_{31}^2=3Q^2$ is employed. The results are given in Table~\ref{tab1}.

\begin{table}[H]
\centering
\caption{\label{tab1}Parameter values estimated using Eq.~(\ref{eq1}) by LHCb Collaboration~\cite{LHCb:2025jci,LHCb:2017pnz} for two- and three-pion Bose--Einstein correlations (BECs). Note that when the low-activity case used the linear long-range correlation term, ${\rm LRC_{(linear,\,3\pi)}}=[C(1+\delta Q)]^{3/2}$, $p_c$ changed from 0.08 to 0.03 and $\chi^2$ changed from 170 to 173. All other values remained unchanged. Multiplicity intervals $N_{\rm ch}^{\rm VELO}\in$ [5, 10], [11, 20] and [21, 60] correspond to $N_{\rm ch}$ intervals [8, 18], [19, 35] and [36, 96], respectively.}
\begin{tabular}{ccccccc}
\hline
\multicolumn{2}{l}{Two-pion BEC}\\
activity, $N_{\rm ch}$ & $R$ (fm) & \multicolumn{2}{c}{$\lambda$} & $C$ & $\delta$ (GeV$^{-1}$) & $\chi^2$/dof\\
\hline
low, [8, 18]
& $1.01\pm 0.10$
& \multicolumn{2}{c}{$0.72\pm 0.05$}
& $0.888\pm 0.002$
& $0.09\pm 0.04$
& 591/386\\

$\!\!\!\!$medium, [19, 35]
& $1.48\pm 0.17$
& \multicolumn{2}{c}{$0.63\pm 0.05$}
& $0.942\pm 0.001$
& $0.05\pm 0.01$
&  623/386\\

high, [36, 96]
& $1.80\pm 0.16$
& \multicolumn{2}{c}{$0.57\pm 0.03$}
& $0.967\pm 0.001$
& $0.03\pm 0.01$
&  621/386\\

\hline\hline
\multicolumn{2}{l}{Three-pion BEC}\\
$N_{\rm ch}$ & $R$ (fm)(inputs) & $f_c$ & $p_c$ & $C$ & $\delta$ (GeV$^{-1}$) & $\chi^2$/dof\\
\hline
[8, 18]
& $1.01$
& $0.85\pm 0.13$
& $0.08\pm 0.01$
& $0.74\pm 0.01$
& $0.07\pm 0.01$
&  170/191\\

[19, 35]
& $1.48$
& $0.83\pm 0.09$
& $0.29\pm 0.04$
& $0.85\pm 0.01$
& $0.04\pm 0.01$
&  285/191\\

[36, 96]
& $1.80$
& $0.81\pm 0.10$
& $0.35\pm 0.04$
& $0.90\pm 0.01$
& $0.03\pm 0.01$
&  262/191\\

\hline
\end{tabular}
\end{table}

Conversely, combining the GL formula from Refs.~\cite{Biyajima:1982un,Biyajima:1990ku} and the dipole expression for the geometric picture, we obtain  the following equation for the two-pion BEC:
\begin{eqnarray}
\left.\frac{N^{(2\pi)}}{N^{\rm BG}}\right|_{\rm GL} = \left[1+ \frac{p^2}{(1+(RQ)^2)^{2}} + \frac{2p(1-p)}{(1+(RQ)^2)}\right]\times {\rm LRC_{(Gauss)}},
\label{eq3}
\end{eqnarray}
where ${\rm LRC_{(Gauss)}}=C/(1+\alpha e^{-\beta Q^2})$~\cite{CMS:2019fur,Mizoguchi:2023zfu}. Concerning the dipole description, the main difference between studies on BEC and the scattering of an electron off a free proton is attributed to the following: The former is related to the range of pion production region, while the latter showed the proton radius $\sqrt{\langle r^2\rangle}=2\sqrt 3R = 0.81$ fm in the dipole description~\cite{Halzen:1984,Greiner:2002bv,Bogoliubov:1959aa}. Moreover, the Breit frame, where there was no energy transfer to the proton was used as the basic calculation~\cite{Halzen:1984,Greiner:2002bv}. For the three-pion case~\cite{Biyajima:1982un,Biyajima:1990ku,Alt:1998nr,Mizoguchi:2000km,Biyajima:2011rbd}, we obtain the following equation:
\begin{eqnarray}
\left.\frac{N^{(3\pi)}}{N^{\rm BG}}\right|_{\rm GL} = \left[1 + \frac{3p^2}{(1+(RQ)^2)^{2}} + \frac{6p(1-p)}{1+(RQ)^2} + \frac{2p^3}{(1+(RQ)^2)^{3}} + \frac{6p^2(1-p)}{(1+(RQ)^2)^2}\right]
\times \left({\rm LRC_{(Gauss)}}\right)^{1.5}.
\label{eq4}
\end{eqnarray}
The parameter values in Table~\ref{tab2} were estimated by employing Eqs.~(\ref{eq3}) and (\ref{eq4}) in the data analysis~\cite{LHCb:2025jci,LHCb:2017pnz} of two- and three-pion BEC.

\begin{table}[H]
\centering
\caption{\label{tab2}Estimated parameters of two- and three-pion Bose--Einstein correlation (BEC). The p-value is defined in Ref~\cite{Bevington:1992aa}.}
\begin{tabular}{cccccc}
\hline
\multicolumn{2}{l}{Two-pion BEC}\\
activity, $N_{\rm ch}$ & $R$ (fm) & $p$ & $C$ & $\chi^2$/dof & p-value (\%)\\
\hline
low, [8, 18]
& $0.99 \pm 0.02 $
& $0.33 \pm 0.01 $
& $1.175\pm 0.033$
&  627/385
& $7\times 10^{-12}$\\

$\!\!\!\!$medium, [19, 35]
& $1.42 \pm 0.02 $
& $0.30 \pm 0.01 $
& $1.039\pm 0.003$
&  533/385
& $8\times 10^{-5}$\\

high, [36, 96]
& $1.80 \pm 0.04 $
& $0.29 \pm 0.01 $
& $1.016\pm 0.002$
&  483/385
& 0.05\\

\hline\hline
\multicolumn{2}{l}{Three-pion BEC}\\
$N_{\rm ch}$ & $R$ (fm) & $p$ & $C$ & $\chi^2$/dof & p-value (\%)\\
\hline
[8, 18]
& $1.07 \pm 0.03 $
& $0.40 \pm 0.01 $
& $1.324\pm 0.152$
&  154/190
& 97\\

[19, 35]
& $1.50 \pm 0.04 $
& $0.36 \pm 0.01 $
& $1.072\pm 0.008$
&  207/190
& 19\\

[36, 96]
& $1.92 \pm 0.06 $
& $0.34 \pm 0.02 $
& $1.042\pm 0.004$
&  154/190
& 97\\

\hline
\end{tabular}
\end{table}

As presented in Table~\ref{tab2}, the estimated values for $R$ (fm) in the two-pion case have almost the same magnitudes as those from the LHCb Collaboration~\cite{LHCb:2017pnz}. The three-pion case also gives very similar values, provided that the dipole expression is assumed for the exchange function ($E_{\rm 2B}$) (see Appendix~\ref{secA}): $\Delta R=R(3\pi)-R(2\pi)=$ 0.08, 0.08, and 0.12 fm, respectively. However, the $\chi^2$ values are still large for two-pion BEC, warranting further consideration.

In this study, we change our standpoint from a single-source picture~\cite{Biyajima:1990ku} to use the GL formula with a two-component picture~\cite{Mizoguchi:2020gnm,Mizoguchi:2023zfu}. We present our new formulas and analysis of two- and three-pion BEC data from the LHCb Collaboration in Section~\ref{sec2}, comparison of two kinds of long-range correlations (LRCs) in Section~\ref{sec3}, and concluding remarks in Section~\ref{sec4}. Appendix~\ref{secA} contains various BE exchange functions and density functions~\cite{Shimoda:1992gb}, while Appendix~\ref{secB} presents an alternative formula. In Appendix~\ref{secC}, we present the degree of coherences for four- and five-pion BECs within the GL formula and the core/halo model. Furthermore, we examine the correspondence between the GL formula and the core/halo model and derive theoretical expressions for $N^{(4\pi)}/N^{\rm BG}$ and  $N^{(5\pi)}/N^{\rm BG}$.

\section{\label{sec2}BEC analysis by means of new two-component GL formula}

We primarily assume a dipole expression for the pion exchange functions $E_{\rm 2B}$ because the electronic form factor of a proton is described by a dipole~\cite{Halzen:1984,Greiner:2002bv}. This hypothesis forms the basis for our interpretation of the parameter $R$. For the mixed term $2p(1-p)$, we assume a weaker inverse power $1/(1+(R_2Q)^2)^{1.5}$ to examine the effect of the coherent component ($1-p$). Comparison of parameters in the core/halo model and the GL formula with geometrical two-component is shown in Table~\ref{tab3}

\begin{table}[H]
\centering
\caption{\label{tab3}Comparison of core/halo model with GL formula with geometrical two-component.}
\begin{tabular}{lll}
\hline
& core/halo model & GL formula with geometrical two-component\\
\hline
multiplicities
& $\langle n_{\rm tot}\rangle = \langle n_{\rm core}\rangle+\langle n_{\rm halo}\rangle$
& $\langle n_{\rm tot}\rangle = A({\rm chaotic})+|\zeta|^2({\rm coherent})$\vspace{2mm}\\

fraction of core
& $f_c=\langle n_{\rm core}\rangle/\langle n_{\rm tot}\rangle$
& \vspace{2mm}\\

chaotic and coherent 
& $p_c= |\zeta_c|^2/\langle n_{\rm core}\rangle$
& ratios $p=A/\langle n_{\rm tot}\rangle$, \\

components in core
& 
& \qquad\ \ $(1-p)=|\zeta|^2/\langle n_{\rm tot}\rangle$\vspace{2mm}\\

degree of coherence
& $\lambda = f_c^2[(1-p_c)^2+2(1-p_c)p_c]$
& $\lambda = p^2+2p(1-p)$\\

in $2\pi$ BEC\vspace{1mm}\\

Bose exchange
& $E_{\rm 2B}=e^{-R_{\rm core}Q}$ ($R=R_{\rm core}$),
& $E_{\rm 2B}=\dfrac{1}{(1+(R_1Q)^2)^2}$ for $p^2$,\\

\ra{function}
& $E_{\rm 2B}=e^{-R_{\rm halo}Q}$ ($R_{\rm halo}\to \infty$)
& $E_{\rm 2B}=\dfrac{1}{(1+(R_2Q)^2)^{1.5}}$ for $2p(1-p)$\vspace{2mm}\\

$2\pi$ BEC formula
& $1+\lambda e^{-RQ}$
& $1+\dfrac{p^2}{(1+(R_1Q)^2)^2}+\dfrac{2p(1-p)}{(1+(R_2Q)^2)^{1.5}}$\\

\hline
physical picture
& Fig.~\ref{fig1}
& Fig.~\ref{fig2}\\

\hline
\end{tabular}
\end{table}

\begin{figure}[H]
  \centering
  \includegraphics[width=0.48\columnwidth]{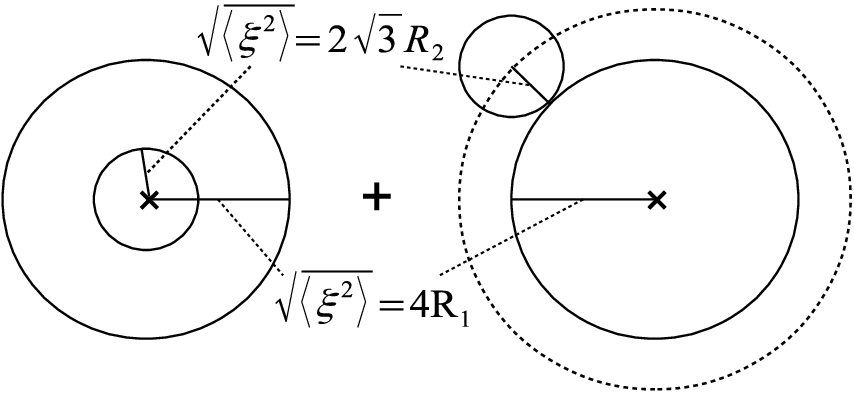}
  \caption{\label{fig2}Schematic mixing description of the Glauber--Lachs (GL) formula based on coalescence and satellite model. The weight factor for the larger region ($R_1$) is $p^2$.}
\end{figure}

\begin{figure}[H]
  \centering
  \includegraphics[width=0.48\columnwidth]{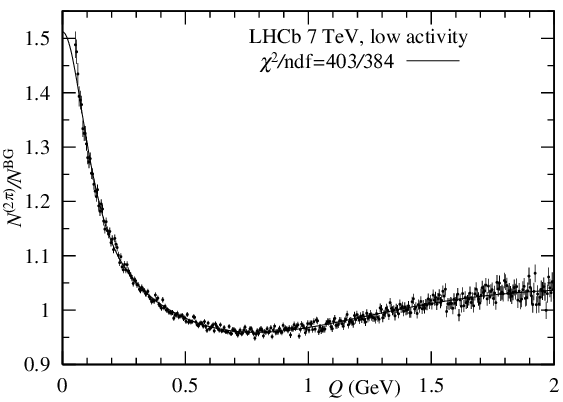}
  \includegraphics[width=0.48\columnwidth]{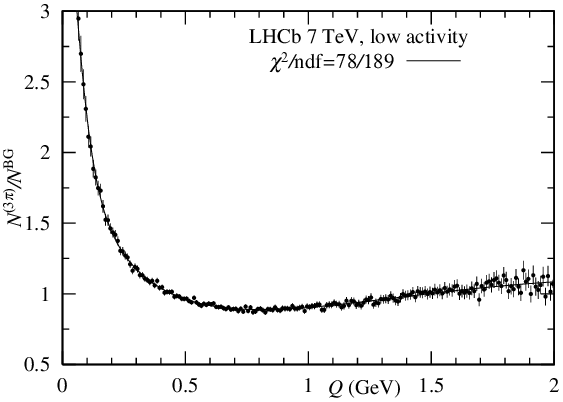}\\
  \includegraphics[width=0.48\columnwidth]{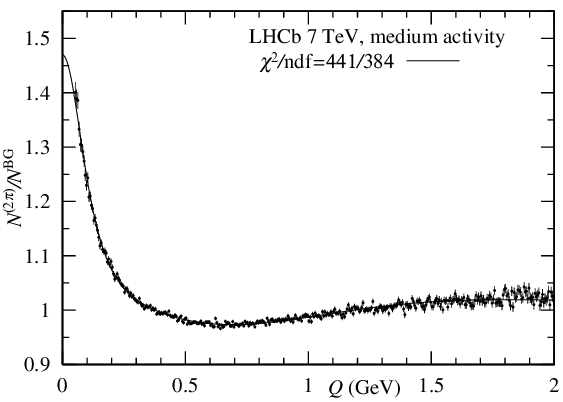}
  \includegraphics[width=0.48\columnwidth]{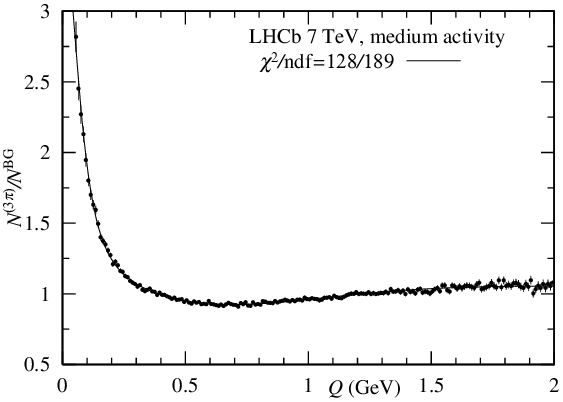}\\
  \includegraphics[width=0.48\columnwidth]{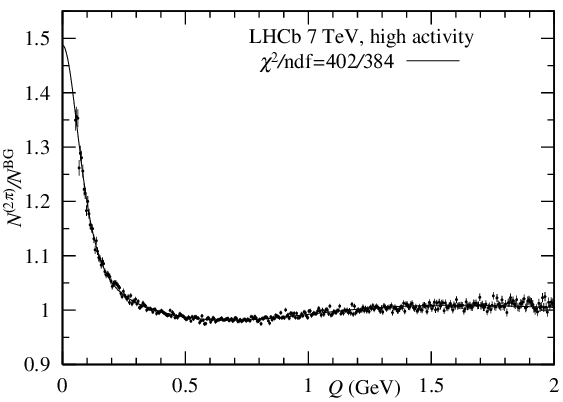}
  \includegraphics[width=0.48\columnwidth]{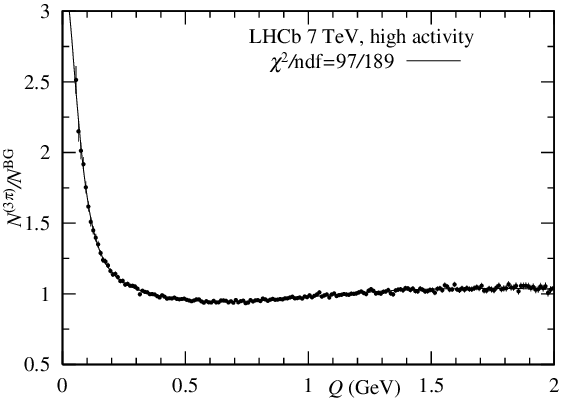}\\
  \caption{\label{fig3}Analysis of two- and three-pion Bose--Einstein correlation by Eqs.~(\ref{eq5}) and (\ref{eq6}).}
\end{figure}

This yields
\begin{eqnarray}
\left.\frac{N^{(2\pi)}}{N^{\rm BG}}\right|_{\rm GL} = \left[1+ \frac{p^2}{(1+(R_1Q)^2)^{2}} + \frac{2p(1-p)}{(1+(R_2Q)^2)^{1.5}}\right]\times {\rm LRC_{(Gauss)}},
\label{eq5}
\end{eqnarray}
\begin{eqnarray}
\left.\frac{N^{(3\pi)}}{N^{\rm BG}}\right|_{\rm GL} &\!\!=&\!\! \left[1 + \frac{3p^2}{(1+(R_1Q)^2)^{2}} + \frac{6p(1-p)}{(1+(R_2Q)^2)^{1.5}} + \frac{2p^3}{(1+(R_1Q)^2)^{3}} + \frac{3p}{(1+(R_1Q)^2)}\frac{2p(1-p)}{(1+(R_2Q)^2)^{1.5}}\right]\nonumber\\
&\!\!&\!\! \times \left({\rm LRC_{(Gauss)}}\right)^{1.5}.
\label{eq6}
\end{eqnarray}
The results obtained from Eqs.~(\ref{eq5}) and (\ref{eq6}) are presented in Table~\ref{tab4} and Fig.~\ref{fig3}.

\begin{table}[H]
\centering
\caption{\label{tab4}Estimated parameters of two- and three-pion Bose--Einstein correlation (BEC) from the two-component Glauber--Lachs formula. The p-values~\cite{Bevington:1992aa} for low-, medium-, and high-activity two-pion BEC are 24 \%, 2.3 \%, and 25 \%, and for three-pion BEC, they are 100 \%, 100 \%, and 100 \%, respectively.}
\begin{tabular}{cccccccc}
\hline
\multicolumn{8}{l}{Two-pion BEC: From top to bottom $N_{\rm ch}$ $[8,\,18]$, $[19,\,35]$, and $[36,\,96]$.}\\
activity & $p$ & $p^2$ & $R_1$ (fm)& $2p(1-p)$ & $R_2$ (fm) & $C$ & $\chi^2$/dof\\
\hline
low
& $0.75$
& $0.56 \pm 0.01$   
& $1.17 \pm 0.05 $
& $0.38 \pm 0.02$
& $0.29 \pm 0.01 $
& $1.036\pm 0.003$
&  403/384\\

$\!\!\!\!$medium
& $0.77$
& $0.59 \pm 0.01$   
& $1.28 \pm 0.04 $
& $0.36 \pm 0.02$
& $0.25 \pm 0.01 $
& $1.004\pm 0.002$
&  441/384\\

high
& $0.78$
& $0.61 \pm 0.02$   
& $1.55 \pm 0.06 $
& $0.34 \pm 0.03$
& $0.24 \pm 0.01 $
& $0.988\pm 0.002$
&  402/384\\

\hline
\multicolumn{8}{l}{From top to bottom column, two sets of $\alpha$ and $\beta$ parameters ($\alpha$, $\beta$ (GeV$^{-2}$)) are}\\ 
\multicolumn{8}{l}{($0.33\pm 0.01$, $0.83\pm 0.02$), ($0.33\pm 0.01$, $1.15\pm 0.03$), and ($0.30\pm 0.01$, $1.28\pm 0.05$).}\\

\hline\hline
\multicolumn{8}{l}{Three-pion BEC}\\
$N_{\rm ch}$ & $p$ & $p^2$ & $R_1$ (fm)& $2p(1-p)$ & $R_2$ (fm) & $C$ & $\chi^2$/dof\\
\hline
[8, 18]
& $0.80$
& $0.64 \pm 0.02$   
& $1.56 \pm 0.15 $
& $0.32 \pm 0.04$
& $0.42 \pm 0.05 $
& $1.083\pm 0.021$
&  78/189\\

[19, 35]
& $0.84$
& $0.71 \pm 0.02$   
& $1.68 \pm 0.11 $
& $0.26 \pm 0.04$
& $0.34 \pm 0.03 $
& $1.028\pm 0.005$
&  128/189\\

[36, 96]
& $0.84$
& $0.71 \pm 0.03$   & $0.27 \pm 0.05$
& $1.80 \pm 0.09 $

& $0.27 \pm 0.02 $
& $1.004\pm 0.005$
&  97/189\\

\hline
\multicolumn{8}{l}{$\!\!\!(\alpha$, $\beta$(GeV$^{-2}$)) are ($0.40\pm 0.04$, $0.62\pm 0.10$), ($0.42\pm 0.05$, $1.18\pm 0.04$) and ($0.47\pm 0.04$, $1.27\pm 0.06$).}\\

\hline\hline
\multicolumn{8}{l}{Three-pion BEC where $R_1$ (fm) and $p$ from the two-pion BEC cases are used as inputs for analysis.}\\
\hline
$N_{\rm ch}$ & $p$ fixed & $p^2$ & $R_1$ (fm) fixed & $2p(1-p)$ & $R_2$ (fm) & $C$ & $\chi^2$/dof\\
\hline
[8, 18]
& $0.75$
& $0.56$
& $1.17$
& $0.38$
& $0.32\pm 0.01$
& $1.046\pm 0.009$
&  100/191\\

[19, 35]
& $0.77$
& $0.59$
& $1.28$
& $0.35$
& $0.25\pm 0.01$
& $1.007\pm 0.003$
&  173/191\\

[36, 96]
& $0.78$
& $0.61$
& $1.55$
& $0.34$
& $0.23\pm 0.01$
& $0.985\pm 0.002$
&  113/191\\

\hline
\end{tabular}
\end{table}

Comparing $R_1$ and $R_2$ for two- and three-pion BEC, we observe that $\Delta R_1=R_1(3\pi)-R_1(2\pi)=$ 0.39, 0.40, and 0.25 fm, and $\Delta R_2=R_2(3\pi)-R_2(2\pi)=$ 0.13, 0.09, and 0.03 fm, The $\Delta R_1$ values in Table~\ref{tab4} are larger than those presented in Table~\ref{tab2}. However, the $\chi^2$ value is smaller in Table~\ref{tab4} than in Table~\ref{tab2}.

When $R_1$ and $p$ from the two-pion BEC analysis (Table~\ref{tab4}) are used as inputs for the three-pion BEC analysis, the resulting $R_2(3\pi)$ is nearly identical to $R_2(2\pi)$. 

By employing $R_1$ and $R_2$, we estimate the extensions in the configuration space $\sqrt{\langle\xi^2\rangle}$ in Table~\ref{tab10} and Appendix~\ref{secA}. We estimate the extensions in the configuration space as the proton electric form factor $\sqrt{\langle r^2\rangle}=0.81$ fm. For larger extensions $R_1$ we obtain $\sqrt{\langle\xi^2\rangle}=4R_1$, while for smaller extensions $\sqrt{\langle\xi^2\rangle}=2\sqrt 3R_2\approx 3.5R_2$ (Table~\ref{tab5}). Therefore, we consider the two-component geometric picture in Fig.~\ref{fig2}, as also proposed by Martin et al.~\cite{Khoze:2016hns}.

\begin{table}[H]
\centering
\caption{\label{tab5}Estimation of extensions in the configuration space.}
\begin{tabular}{ccc}
\hline
activity & $\sqrt{\langle\xi^2\rangle}$ (fm) for $R_1$ & $\sqrt{\langle\xi^2\rangle}$ (fm) for $R_2$\\
\hline
low     & $4.68\pm 0.20$  & $1.00\pm 0.09$\\
medium  & $5.12\pm 0.16$  & $0.87\pm 0.09$\\
high    & $6.20\pm 0.24$  & $0.83\pm 0.09$\\
\hline
\end{tabular}
\end{table}

The magnitudes of the parameters in Table~\ref{tab4} are estimated in the configuration space and summarized in Table~\ref{tab5}. Moreover, $f_c=\langle n_{\rm core}\rangle/\langle n_{\rm tot}\rangle$ in Table~\ref{tab1} is compared with $p(2\pi)$ and $p(3\pi)$ in Table~\ref{tab4}, as shown in Table~\ref{tab6}.

\begin{table}[H]
\centering
\caption{\label{tab6}Comparison of two parameters $f_c=\langle n_{\rm core}\rangle/\langle n_{\rm tot}\rangle$ in the core/halo model from Table~\ref{tab1} and $p=A({\rm chaotic})/\langle n_{\rm tot}\rangle$ for $p(2\pi)$ and $p(3\pi)$ from Table~\ref{tab4}.}
\begin{tabular}{c|c|c}
\hline
activity & core/halo model & GL formula with two-component model\\
\hline
& $f_c=\langle n_{\rm core}\rangle/\langle n_{\rm tot}\rangle$
& $p=A({\rm chaotic})/\langle n_{\rm tot}\rangle$, 
 $(p_{2\pi}+p_{3\pi})/2$\\
low
& $0.83\pm 0.13$
& $(0.75+0.80)/2$\qquad $0.78\pm 0.01$\\

medium
& $0.83\pm 0.09$
& $(0.77+0.84)/2$\qquad $0.81\pm 0.01$\\

high
& $0.81\pm 0.10$
& $(0.78+0.84)/2$\qquad $0.81\pm 0.01$\\
\hline
\end{tabular}
\end{table}

They probably correspond to categories of the collision mechanism. Our previous data analysis at 13 TeV from the CMS Collaboration also supports a similar picture~\cite{CMS:2019fur,Mizoguchi:2023zfu}. Therefore, we expect the following correspondences for the LHC energies:
\begin{eqnarray*}
\left\{
\begin{array}{l}
\mbox{larger region ($R_1$) $\cdots$ pure chaotic term ($p^2$)}\\
\mbox{smaller region ($R_2$) $\cdots$ mixed term of chaotic and coherent components ($2p(1-p)$)}
\end{array}
\right.
\end{eqnarray*}

In conclusion, the GL formula with a dipole form for the chaotic part and an inverse one-and-a-half pole for the mixed term $2p(1-p)=A|\zeta|^2/\langle n_{\rm tot}\rangle^2$ works fairly well. 

Our analysis shows that the $R_2$ values are very small, corresponding to localized spots in the large-range $R_1$'s. Indeed, the same charged pions are produced from a given spot, the phase difference between two pions seems to be very small, probably reflecting coherent pion production. On the other hand, the larger $\sqrt{\langle\xi^2\rangle}\approx 5$ fm region reflects the decay lengths of resonances produced at the LHC: For example $c\tau_{\rho}=1.3$ fm for the $\rho$-meson, $c\tau_{\omega}=23$ fm for the $\omega$-meson and $c\tau_{\Delta(1232)}=1.7$ fm for the $\Delta(1232)$-baryon, respectively~\cite{Grishin:1971wu,Grassberger:1976au,Lednicky:1992me,Bolz:1992hc}.

\section{\label{sec3}Comparison of two types of LRCs}

In Table \ref{tab4}, we used a Gaussian form for the LRC.  In this study, we investigate linear forms, ${\rm LRC_{(linear)}}=C(1+\delta Q)$, and  ${\rm LRC_{(linear,\,3\pi)}}=[C(1+\delta Q)]^{3/2}$ for three-pion BEC. 

\begin{figure}[H]
  \centering
  \includegraphics[width=0.48\columnwidth]{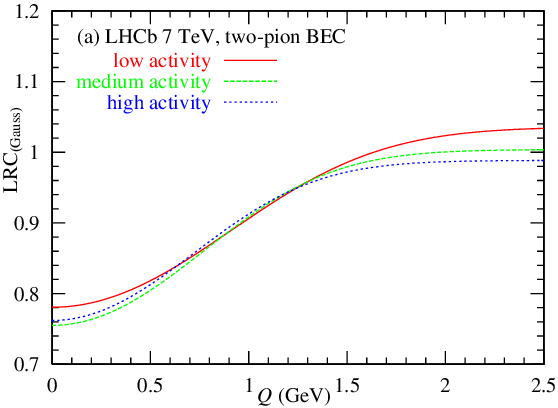}
  \includegraphics[width=0.48\columnwidth]{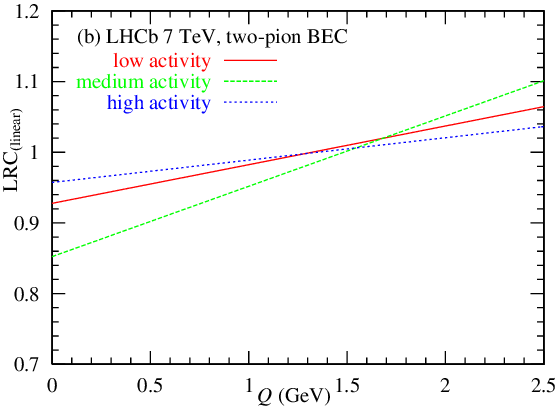}\\
  \includegraphics[width=0.48\columnwidth]{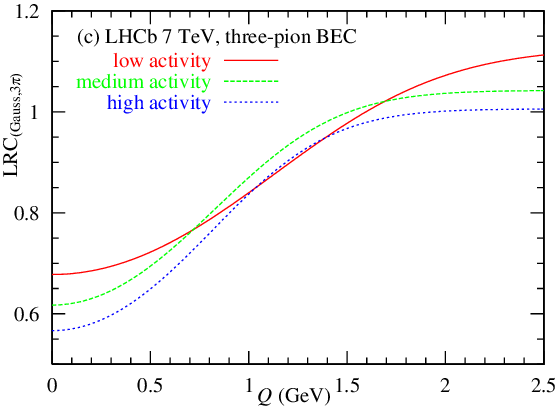}
  \includegraphics[width=0.48\columnwidth]{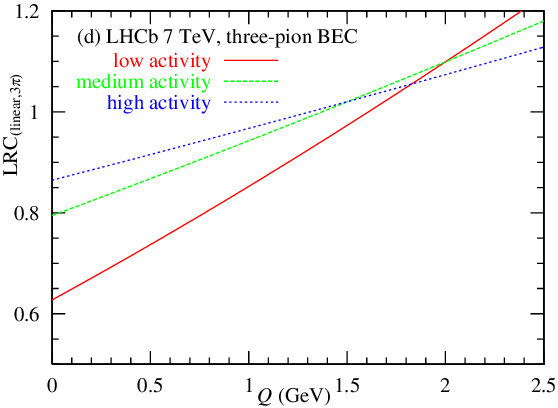}
  \caption{\label{fig4}Comparison of the two types of long-range correlations obtained using Eqs.~(\ref{eq5}) and (\ref{eq6}). (a) ${\rm LRC_{(Gauss)}}=C/(1+\alpha e^{-\beta Q^2})$, (b) ${\rm LRC_{(linear)}}=C(1+\delta Q)$, (c) ${\rm LRC_{(Gauss,\,3\pi)}}=\left[C/(1+\alpha e^{-\beta Q^2})\right]^{1.5}$, and (d) ${\rm LRC_{(linear,\,3\pi)}}=[C(1+\delta Q)]^{3/2}$.}
\end{figure}

A notable point in the results (Table~\ref{tab7}) is that the estimated normalization factors $C$ are less than 1.0. In other words, the linear LRCs increase between $Q=0.0$ and 2.0 GeV. Moreover, for three-pion BEC, the $\chi^2$ values in Table~\ref{tab7} are larger than those in Table~\ref{tab4}. In Fig.~\ref{fig4}, we compare the two types of LRCs.

As presented in Tables~\ref{tab4} and \ref{tab7}, the estimated parameters are LRC-dependent; therefore, we need to simultaneously consider the LRC and geometrical expressions.

\begin{table}[H]
\centering
\caption{\label{tab7}Estimated parameters of two- and three-pion Bose--Einstein correlation (BEC) obtained from Eqs.~(\ref{eq5}) and (\ref{eq6}) with ${\rm LRC_{(linear)}}=C(1+\delta Q)$ /or ${\rm LRC_{(linear,\,3\pi)}}=[C(1+\delta Q)]^{3/2}$.  The p-values for the three activity classes of two-pion BEC are 3.0 \%, $1.3\times 10^{-5}$ \%, and $3.9\times 10^{-5}$ \%, and for three-pion BEC, they are 100 \%, 55 \%, and 55 \%, respectively.}
\begin{tabular}{ccccccc}
\hline
\multicolumn{2}{l}{Two-pion BEC}\\
activity, $N_{\rm ch}$ & $R_1$ (fm) & $R_2$ (fm) & $p$ & $C$ & $\delta$ (GeV$^{-1}$) & $\chi^2$/dof \\ 
\hline
low, [8, 18]
& $1.80 \pm 0.05 $
& $0.50 \pm 0.01$
& $0.76 \pm 0.01$
& $0.852\pm 0.004$
& $0.117\pm 0.004$
&  439/385\\

$\!\!\!\!$medium, [19, 35]
& $2.87 \pm 0.08 $
& $0.81 \pm 0.02$
& $0.82 \pm 0.01$
& $0.928\pm 0.001$
& $0.059\pm 0.001$
&  545/385\\

high, [36, 96]
& $3.23 \pm 0.11 $
& $0.95 \pm 0.03$
& $0.86 \pm 0.01$
& $0.957\pm 0.001$
& $0.033\pm 0.001$
&  538/385\\

\hline\hline
\multicolumn{2}{l}{Three-pion BEC}\\
$N_{\rm ch}$ & $R_1$ (fm) & $R_2$ (fm) & $p$ & $C$ & $\delta$ (GeV$^{-1}$) & $\chi^2$/dof\\ 
\hline
[8, 18]
& $1.47 \pm 0.08 $
& $0.47 \pm 0.03$
& $0.77 \pm 0.02$
& $0.733\pm 0.014$
& $0.227\pm 0.016$
&  92/190\\

 [19, 35]
& $2.49 \pm 0.09 $
& $0.74 \pm 0.03$
& $0.82 \pm 0.01$
& $0.858\pm 0.004$
& $0.121\pm 0.004$
&  187/190\\

[36, 96]
& $3.05 \pm 0.13 $
& $0.91 \pm 0.05$
& $0.87 \pm 0.02$
& $0.908\pm 0.003$
& $0.078\pm 0.003$
&  187/190\\

\hline\hline
\end{tabular}
\end{table}

\section{\label{sec4}Concluding remarks}
The $\chi^2/$ndf values in Table~\ref{tab1} may suggest that the core/halo model is reflecting to idealized hypothesis, because the halo range $R_h\to \infty$ and the two terms in $\lambda_2$ in Eq.~(\ref{eq1}) are treated identical to $e^{-RQ}$: $f_c^2(1-p_c)^2e^{-RQ}$ and $f_c^22p_c(1-p_c)e^{-RQ}$. On the other hand, as shown in Table~\ref{tab2}, the GL formula proposed in 1990~\cite{Biyajima:1990ku} is necessary some improvement to describe the BEC data from the LHCb Collaboration.

To explain the situation above mentioned, more concretely, we present results by the following conventional formula~\cite{Mizoguchi:2023zfu} in Table~\ref{tab8},
\begin{eqnarray}
\left.\frac{N^{(2\pi)}}{N^{\rm BG}}\right|_{\rm conv.\ formula} = \left[1 + \lambda_1 e^{-R_1Q} + \lambda_2 e^{-(R_2Q)^2}\right]\times {\rm LRC_{(Gauss)}},
\label{eq7}
\end{eqnarray}
where two degree of coherences $\lambda_1$ and $\lambda_2$ are less than 1.0: The constraints $(\lambda_1\ {\rm and}\ \lambda_2)\le 1.0$ and $\lambda_1 + \lambda_2\le 1.0$ are used in concrete analysis.

\begin{table}[H]
\centering
\caption{\label{tab8}Estimated parameters of two-pion Bose--Einstein correlation (BEC) by Eq.~(\ref{eq7}).}
\begin{tabular}{ccccccc}
\hline
\multicolumn{2}{l}{Two-pion BEC}\\
activity, $N_{\rm ch}$ & $R_1$ (fm) & $R_2$ (fm) & $\lambda_1$ & $\lambda_2$ & $C$ & $\chi^2$/dof \\
\hline
low, [8, 18]
& $1.93\pm 0.11$
& $0.37\pm 0.02$
& $0.77\pm 0.02$
& $0.17\pm 0.01$
& $1.056\pm 0.009$
& 383/383\\

$\!\!\!\!$medium, [19, 35]
& $2.52\pm 0.11$
& $0.46\pm 0.02$
& $0.78\pm 0.02$
& $0.10\pm 0.01$
& $1.028\pm 0.002$
& 392/383\\

high, [36, 96]
& $3.04\pm 0.13$
& $0.44\pm 0.03$
& $0.82\pm 0.03$
& $0.10\pm 0.01$
& $1.010\pm 0.001$
& 369/383\\

\hline
\multicolumn{7}{l}{$\!\!\!(\alpha,\ \beta$(GeV$^{-2}$)) are ($0.18\pm 0.01$, $0.62\pm 0.11$), ($0.10\pm 0.01$, $0.84\pm 0.09$) and ($0.07\pm 0.01$, $1.39\pm 0.20$).}\\

\hline
\end{tabular}
\end{table}

To address the problems mentioned above, and following the two-component model proposed by Khoze et al.~\cite{Khoze:2016hns}, we propose the GL formula with a two-component geometrical model as Eqs.~(\ref{eq5}) and (\ref{eq6}). For the first term with $p^2$, we chose the dipole expression, and for the second term $2p(1-p)$, we assumed a weaker expression, $1/(1+(RQ)^2)^{1.5}$, than the dipole form. The physical picture underlying the present paper are shown in Fig.~\ref{fig2}. At present, it is difficult to distinguish between the coalescense and the satellite models.

\begin{figure}[H]
  \centering
  \includegraphics[width=0.48\columnwidth]{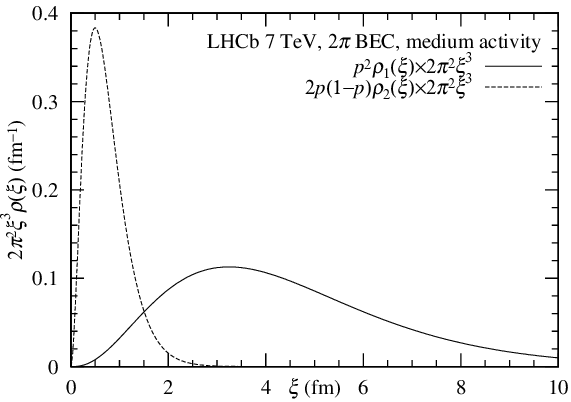}
  \includegraphics[width=0.48\columnwidth]{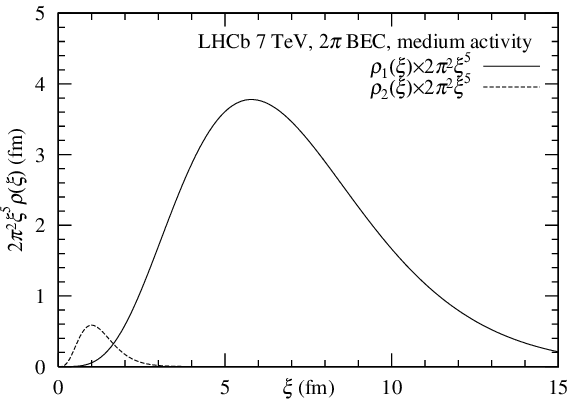}
  \caption{\label{fig5}Density functions types and average square extension ($\langle \xi^2\rangle$) with modified Bessel functions of medium activity for two-pion Bose--Einstein correlation ($2\pi$ BEC) in Table~\ref{tab4}.}
\end{figure}

\begin{figure}[H]
  \centering
  \includegraphics[width=0.48\columnwidth]{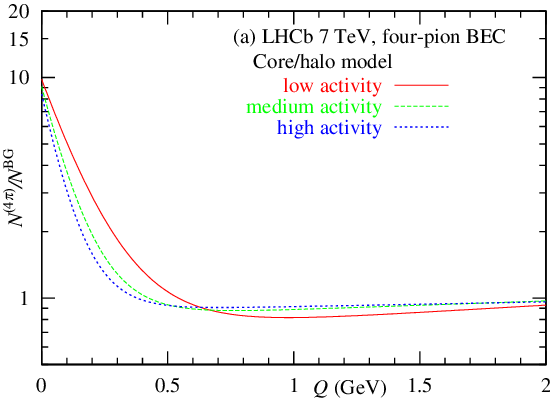}
  \includegraphics[width=0.48\columnwidth]{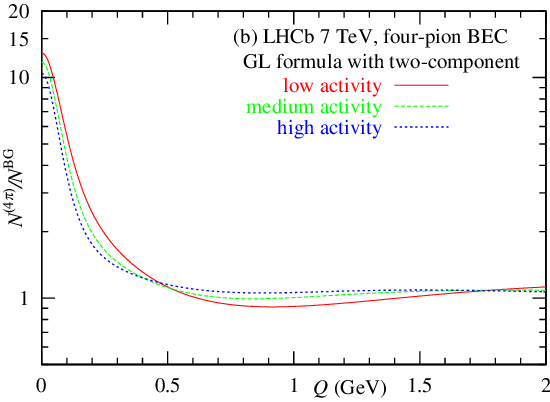}
  \caption{\label{fig6}Predictions of four-pion Bose--Einstein correlation using Eqs.~(\ref{eq10}) and (\ref{eq11}).}
\end{figure}

\noindent
{\bf C1)} The authors of Refs.~\cite{LHCb:2025jci,Greiner:2002bv} applied the core/halo model to analyze their data on two- and three-pion BEC. The results for $f_c=\langle n_{\rm core}\rangle/\langle n_{\rm tot}\rangle$, $f_c^2=0.83\pm 0.13$, $0.83\pm 0.09$ and $0.81\pm 0.10$ are very interesting, because the contribution from the halo component (pions from decayed resonances) is known to be approximately 20 \%. However, we still question on $\chi^2$/ndf values (Table~\ref{tab1}). Moreover, it is difficult to estimate $\sqrt{\langle\xi^2\rangle}$ for the exponential distribution in the configuration space (see Appendix~\ref{secA}).
\bigskip\\
{\bf C2)} We analyzed the two- and three-pion BEC data in $pp$ collisions at 7 TeV from the LHCb Collaboration~\cite{LHCb:2025jci} using a theoretical formula based on quantum optics~\cite{Glauber:1966,Lachs:1965zz} as proposed in \cite{Biyajima:1990ku}. For the geometrical picture, we assumed a dipole expression in four-dimensional space-time. The estimated $R(2\pi)$ values in Table~\ref{tab2} are nearly identical to those reported by the LHCb Collaboration (Table~\ref{tab1}). However, we did not observe any improvements in the $\chi^2$/ndf values (Table~\ref{tab2}).
\bigskip\\
{\bf C3)} We introduced a one-and-a-half pole for the mixed component $2p(1-p)$, which has a weaker power than that of the dipole, to improve the $\chi^2$ values in Table~\ref{tab2}. Specifically, $1/(1+(R_2Q)^2)^{1.5}$ was introduced for the mixed term in $2p(1-p)$. For the coherent component, we assumed the half-pole $1/(1+(R_2Q)^2)^{0.5}$; the results are shown in Table~\ref{tab4}. Moreover, when $R_1(2\pi)$ and $p(2\pi)$ from Table~\ref{tab4} were used as inputs to the working hypothesis of the LHCb Collaboration, we obtained improved $R_2(3\pi)$ and $\chi^2$ values that closely match with the $R_2(2\pi)$ and $\chi^2$ values in Table~\ref{tab1}. In conclusion, the role of the one-and-a-half pole $1/(1+(R_2Q)^2)^{1.5}$ seems to be important in addition to the dipole $1/(1+(R_2Q)^2)^{2.0}$.
\bigskip\\
{\bf C4)} In our analysis, we compared two LRC types: at $Q=2$ GeV, the ${\rm LRC_{(Gauss)}}$ values are close to the normalization factors $C$, which are asymptotic values. Conversely, ${\rm LRC_{(linear)}}$ continues to increase for $Q>2$ GeV.
\bigskip\\
{\bf C5)} We can compare the density functions with estimated parameters $R_1=1.28$ fm, $R_2=0.25$ fm, and $p=0.77$ in Table~\ref{tab4} and Fig.~\ref{fig5} and are computed by using formulas in Table~\ref{tab10}.
\begin{eqnarray*}
\left\{
\begin{array}{l}
p^2\cdot\dfrac{K_{0}(\xi/R)}{8\pi^2R^4}\times 2\pi^2\xi^3\\
2p(1-p)\cdot\dfrac{K_{1/2}(\xi/R)/\sqrt{\xi}}{2^{2.5}\pi^2\Gamma(3/2)R^{3.5}}\times 2\pi^2\xi^3
\end{array}
\right.
\end{eqnarray*}
where $2\pi^2\xi^3$ is the phase space in the four-dimensional Euclidean space, and $K_{0}(\xi/R)$ and $K_{1/2}(\xi/R)$ are modified Bessel functions. The different roles are presented in Fig.~\ref{fig5}.
%
\noindent
{\bf D1)} In Appendix~\ref{secB}, we study the alternative expression for the mixed term $2p(1-p)$, because it is unclear whether the one-and-a-half pole form is optimal or not. The second candidate is the Gaussian distribution. Presently, as shown in Appendix~\ref{secB}, we cannot conclude that the Gaussian distribution is better than the one-and-a-half pole form.
\bigskip\\
{\bf D2)} Using Eqs.~(\ref{eq10}) and (\ref{eq11}) in Appendix~\ref{secC}, we can calculate two forms of $N^{(4\pi)}/N^{\rm BG}$ for empirical analyses by the LHCb Collaboration based on the core/halo model and the GL formula with a geometrical two-component model. These results are shown in Fig.~\ref{fig6}. The maximum values at $Q=0$ GeV are listed in Table~\ref{tab9}. The two approaches may be distinguished clearly if data on four-pion BEC become available.

\begin{table}[H]
\centering
\caption{\label{tab9}The expecting highest values for four-pion BEC at $Q=0$.}
\begin{tabular}{ccc}
\hline
activity & core/halo model & GL formula with two-component\\
\hline
low    & 9.8 & 12.9\\
medium & 9.1 & 11.7\\
high   & 8.4 & 10.5\\
\hline
\end{tabular}
\end{table}

\bigskip\noindent
{\bf D3)} Because of the bunching effect for identical pions under BE statistics, the pion energy differences $(p_{10}-p_{20})$ in each bin are expected to be small. In this idealized case, physical quantities should be interpreted in the coordinate space; for example, $\sqrt{\langle \xi^2\rangle}\to \sqrt{\langle r^2\rangle}$. This situation is similar to the Breit frame.
\bigskip\\
{\it Acknowledgments.} M.B. would like to thank colleagues at the Center for General Education of Shinshu University.

\appendix
\section{\label{secA}Various BE exchange and density functions, and dipole for the proton form factor}

In Table~\ref{tab10}, we present various BE exchange functions $E_{\rm 2B}(Q)$ and density functions $\rho(\xi)$ with the corresponding extension, root-mean-squared of the extension, and fluctuation $\Delta \xi = \sqrt{\langle\xi^2\rangle-\langle\xi\rangle^2}$. $\xi=\sqrt{|\bm r_1-\bm r_2|^2+(t_1-t_2)^2}$. In Table~\ref{tab10}, we omit the index ``E'' for the variable $Q_{\rm E}=\sqrt{|\bm p_1-\bm p_2|^2+(p_{10}-p_{20})^2}$.

\begin{table}[H]
\centering
\caption{\label{tab10}Bose--Einstein exchange function $E_{\rm 2B}(Q)$ in momentum space and the density function $\rho(\xi)$ in configuration space. The average extension $\langle\xi\rangle$, the extension squared $\sqrt{\langle\xi^2\rangle}$ and the fluctuation $\Delta \xi$. $\xi=\sqrt{|\bm r_1-\bm r_2|^2+(t_1-t_2)^2}$ are the corresponding 4-dimensional space-time quantities. Wick rotation and inverse rotation are used if necessary. Note: $K_{2-\kappa}(\xi/R)$ are the modified Bessel functions. $G_E(Q^2)$ is the proton electric form factor~\cite{Halzen:1984}.}
\begin{tabular}{ccccc}
\hline

$E_{\rm 2B}(Q)$ 
& $\rho(\xi)$ 
& $\langle\xi\rangle$ 
& $\sqrt{\langle\xi^2\rangle}$ 
& $\Delta \xi$
\\

\hline

$\dfrac{1}{(1+(RQ)^2)^{\kappa}}$ 
& $\dfrac{\xi^{\kappa-2}K_{2-\kappa}(\xi/R)}{2^{\kappa+1}\pi^2\Gamma(\kappa)R^{\kappa+2}}$ 
\smallskip\\

$\dfrac{1}{(1+(RQ)^2)^{2}}$ 
& $\dfrac{K_{0}(\xi/R)}{8\pi^2R^4}$ 
& $\dfrac 98\pi R$ 
& $4R$
& $1.88R$
\smallskip\\

$\dfrac{1}{(1+(RQ)^2)^{1.5}}$ 
& $\dfrac{K_{1/2}(\xi/R)/\sqrt{\xi}}{2^{2.5}\pi^2\Gamma(3/2)R^{3.5}}$ 
& $3R$ 
& $2\sqrt 3R$
& $\sqrt 3R$
\smallskip\\

$\dfrac{1}{(1+(RQ)^2)^{1.0}}$ 
& $\dfrac{K_{1}(\xi/R)/\xi}{2^{2}\pi^2\Gamma(1.0)R^{3}}$ 
& $\dfrac 34\pi R$ 
& $2\sqrt 2R$
& $1.57R$
\smallskip\\

$e^{-(RQ)^2}$
& $\dfrac{1}{16\pi^2R^4}e^{-\xi^2/(4R^2)}$ 
& $\dfrac 32\sqrt{\pi}R$ 
& $2\sqrt 2R$
& $0.97R$
\smallskip\\

$e^{-RQ}$
& $\dfrac{3/4}{\pi^2R^4}\dfrac{1}{(1+(\xi/R)^2)^{2.5}}$ 
& $\ln \Lambda_{\rm cutoff}/\infty$
& ---
& ---
\\

\hline

\multicolumn{4}{l}{$d=3$ dimension (: proton electric form factor, $G_E(Q^2=-q^2=|\bm q|^2)$).}\\

$G_E(Q^2=-q^2)$ 
& $\rho(r)$ 
& $\langle r\rangle$ 
& $\sqrt{\langle r^2\rangle}$ 
& $\Delta r = \sqrt{\langle r^2\rangle-\langle r\rangle^2}$\\

\hline
$\dfrac{1}{(1+(RQ)^2)^{2}}$ 
& $\dfrac{e^{-r/R}}{8\pi^2R^3}$ 
& $3R$ 
& $2\sqrt 3R$
& $\sqrt 3R$\\

\hline

\end{tabular}
\end{table}

\section{\label{secB}An alternative expression for the mixed term $2p(1-p)$}

\begin{figure}[H]
  \centering
  \includegraphics[width=0.48\columnwidth]{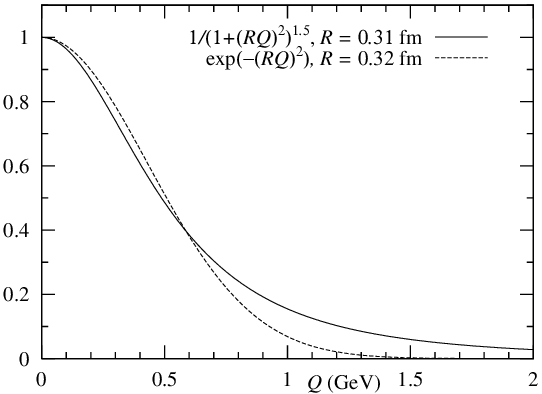}
  \caption{\label{fig7}Comparison of the one-and-a-half pole and a Gaussian distribution.}
\end{figure}

To investigate the role of the one-and-a-half pole form for the mixed term $2p(1-p)$, we propose a Gaussian distribution as an alternative study. The main reason is presented in Fig.~\ref{fig5}. The behaviors of the two kinds of distributions are similar in the region of variable $Q$, where $0<Q<0.7$ GeV (Fig.~\ref{fig7}).

In the data analysis of BEC, we use the following equations:
\begin{eqnarray}
\left.\frac{N^{(2\pi)}}{N^{\rm BG}}\right|_{\rm GL} = \left[1+ \frac{p^2}{(1+(R_1Q)^2)^{2.0}} + 2p(1-p)e^{-(R_2Q)^2}\right]\times {\rm LRC_{(Gauss)}},
\label{eq8}
\end{eqnarray}
\begin{eqnarray}
\left.\frac{N^{(3\pi)}}{N^{\rm BG}}\right|_{\rm GL} &\!\!\!\!\!\!=&\!\! \left[1 + \frac{3p^2}{(1+(R_1Q)^2)^{2.0}} + 6p^2(1-p)e^{-(R_2Q)^2} + \frac{2p^3}{(1+(R_1Q)^2)^{3.0}} + \frac{3p}{1+(R_1Q)^2}2p(1-p)e^{-(R_2Q)^2}\right]\nonumber\\
&\!\!&\!\! \times \left({\rm LRC_{(Gauss)}}\right)^{1.5}.
\label{eq9}
\end{eqnarray}
Table~\ref{tab11} presents the results obtained from using Eqs.~(\ref{eq8}) and (\ref{eq9}). Comparing $\chi^2$ values between Tables~\ref{tab4} and \ref{tab11}, Eqs.~(\ref{eq5}) and (\ref{eq6}) are more favorable than Eqs.~(\ref{eq8}) and (\ref{eq9}); the high activity in Table~\ref{tab11} ($\chi^2=383$) is smaller than that in Table~\ref{tab4} ($\chi^2=402$). This is an exceptional case in the present comparison. In Table~\ref{tab11}, the differences $\Delta R_1=R_1(3\pi)-R_1(2\pi)=$ 0.24, 0.24, and 0.21 fm are for low, medium, and high activities, respectively.

The $\Delta R_1$ values for Table~\ref{tab11} are smaller than those in Table~\ref{tab4}. However, the results of the medium activity are still exceptional. In conclusion, the Gaussian distribution appears to be a worthwhile second candidate for the mixed term.

\begin{table}[H]
\centering
\caption{\label{tab11}Estimated parameters of two- and three-pion Bose--Einstein correlation (BEC) by Eqs.~(\ref{eq8}) and (\ref{eq9}). The p-values for the three activity levels of two-pion BEC are 2.3 \%, 0.9 \%, and 50 \%, and for three-pion BEC, they are 100 \%, 98 \%, and 100 \%, respectively.}
\begin{tabular}{cccccc}
\hline
\multicolumn{2}{l}{Two-pion BEC}\\
activity, $N_{\rm ch}$ & $R_1$ (fm) & $R_2$ (fm) & $p$ & $C$ & $\chi^2$/dof\\
\hline
low, $[8,\,18]$
& $1.00 \pm 0.03 $
& $0.30 \pm 0.01 $
& $0.76 \pm 0.01 $
& $1.035\pm 0.002$
&  441/384\\

$\!\!\!\!$medium, $[19,\,35]$
& $1.20 \pm 0.03 $
& $0.32 \pm 0.01 $
& $0.77 \pm 0.01 $
& $1.020\pm 0.001$
&  452/384\\

high, $[36,\,96]$
& $1.48 \pm 0.04 $
& $0.35 \pm 0.01 $
& $0.77 \pm 0.01 $
& $1.008\pm 0.001$
&  383/384\\

\hline
\multicolumn{6}{l}{$\!\!\!(\alpha,\ \beta$(GeV$^{-2}$)) are ($0.37\pm 0.01,\ 1.23\pm 0.03$), ($0.37\pm 0.01,\ 1.85\pm 0.05$) and ($0.34\pm 0.01$, $2.42\pm 0.08$).}\\

\hline\hline
\multicolumn{2}{l}{Three-pion BEC}\\
$N_{\rm ch}$ & $R_1$ (fm) & $R_2$ (fm) & $p$ & $C$ & $\chi^2$/dof\\
\hline
[8, 18]
& $1.24 \pm 0.08 $
& $0.34 \pm 0.02 $
& $0.84 \pm 0.01 $
& $1.045\pm 0.008$
&  93/189\\

[19, 35]
& $1.44 \pm 0.05 $
& $0.35 \pm 0.01 $
& $0.84 \pm 0.01 $
& $1.033\pm 0.003$
&  150/189\\

[36, 96]
& $1.69 \pm 0.06 $
& $0.35 \pm 0.01 $
& $0.83 \pm 0.01 $
& $1.025\pm 0.002$
&  100/189\\

\hline

\multicolumn{6}{l}{$\!\!\!(\alpha,\ \beta$(GeV$^{-2}$)) are ($0.47\pm 0.05,\ 1.12\pm 0.08$), ($0.54\pm 0.04,\ 1.80\pm 0.07$) and ($0.55\pm 0.04$, $2.14\pm 0.09$).}\\
\hline
\end{tabular}
\end{table}

\section{\label{secC}Description of four- and five-pion BEC by the core/halo model and the GL formula with two-component}

\begin{figure}[H]
  \centering
  \includegraphics[width=0.65\columnwidth]{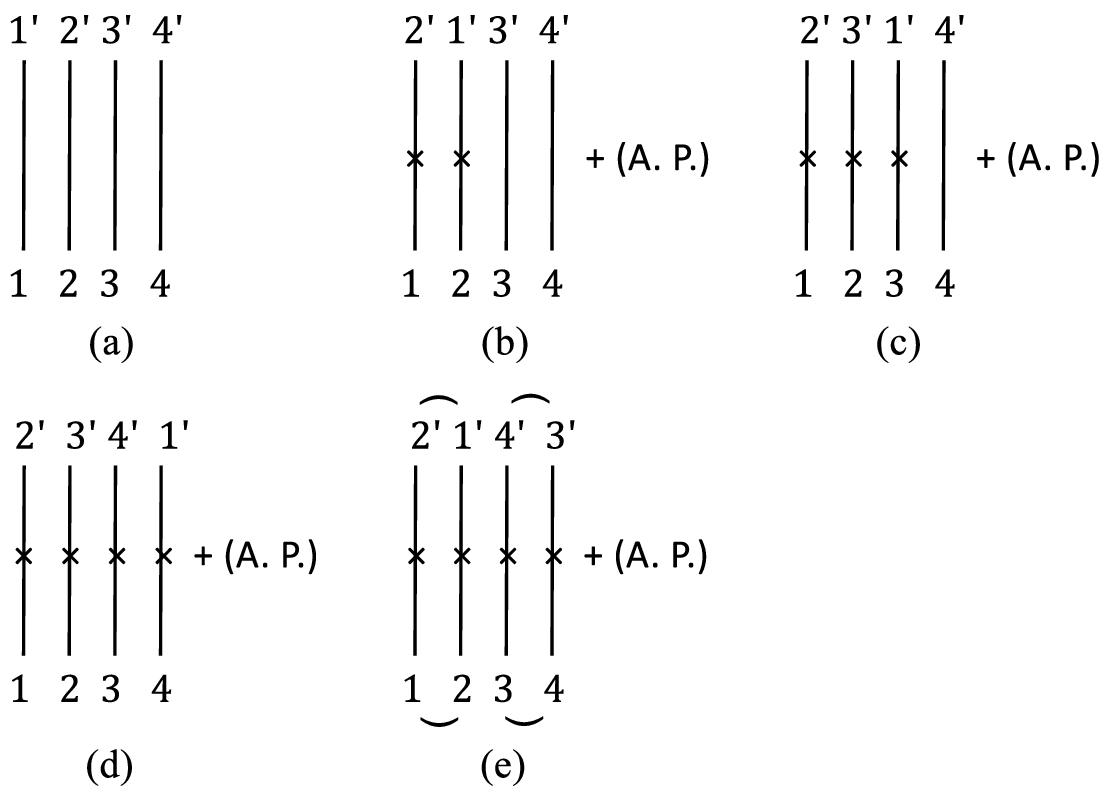}
  \caption{\label{fig8}Diagrammatic picture of the fourth moment $F^{(4-)}$ expression in the GL formula. ``$\times$''  indicate the exchange of identical pions. The exact correspondence between ($1 + \lambda_{4}^{\rm (GL)}$) and ($1 + \lambda_{4}^{\rm (core/halo)}$) does not hold for the last term (e). The  $f_c^2p_c^2$ term corresponding to $(1-p)^2$ in ($1 + \lambda_{4}^{\rm (GL)}$) disappears from ($1 + \lambda_{4}^{\rm (core/halo)}$).}
\end{figure}

The degrees of coherence in the core/halo model and the GL formula are explicitly given in Table~\ref{tab12}. Moreover, the diagrammatic expansion of the fourth moment $F^{(4-)} = \langle n(n-1)(n-2)(n-3)\rangle$ in the GL formula is presented in Fig.~\ref{fig8}. To obtain the corresponding expressions in the core/halo model, replace ``$p$'' and ``$1-p$''in the GL formula with ``$f_c(1-p_c)$'' and ``$f_cp_c$,'' respectively.

\begin{table}[H]
\centering
\caption{\label{tab12}Analytical expressions of four- and five-pion Bose--Einstein correlation (BEC) in the GL formula and core/halo model. Note that the fifth in ($1 + \lambda_{4}^{\rm (GL)}$), and the fifth and the seventh terms in ($1 + \lambda_{5}^{\rm (GL)}$) are partially different from $(1 + \lambda_{4}^{\rm (core/halo)})$ and $(1 + \lambda_{5}^{\rm (core/halo)})$, respectively. In other words, there is no existence of ``$f_c^2p_c^2$'' team in them, because of the hypothesis introduced in the core/halo model.}
\begin{tabular}{cl}
\hline

$F^{(4-)}$ & $(A+|\zeta|^2)^4+6(A+|\zeta|^2)^2(A^2+2A|\zeta|^2)+8(A+|\zeta|^2)(A^3+3A^2|\zeta|^2)$\\
& $+6(A^4+4A^3|\zeta|^2)+3(A^2+2A|\zeta|^2)^2$\\

\hline

$F^{(5-)}$ & $(A+|\zeta|^2)^5+10(A+|\zeta|^2)^3(A^2+2A|\zeta|^2)+20(A+|\zeta|^2)^2(A^3+3A^2|\zeta|^2)$\\
& $+30(A+|\zeta|^2)(A^4+4A^3|\zeta|^2)+15(A+|\zeta|^2)(A^2+2A|\zeta|^2)^2$\\
& $+24(A^5+5A^4|\zeta|^2)+20(A^3+3A^2|\zeta|^2)(A^2+2A|\zeta|^2)$\\

\hline\hline

$1 + \lambda_{4}^{\rm (GL)}$ & $1 + 6[p^2 + 2p(1-p)]+ 8[p^3 + 3p^2(1-p)]$\\
& $+6[p^4 + 4p^3(1-p)] + 3[p^2 + 2p(1-p)]^2$\\

\hline

$1 + \lambda_{5}^{\rm (GL)}$ &
$1 + 10[p^2 + 2p(1-p)] + 20[p^3 + 3p^2(1-p)]$\\
& $+ 30[p^4 + 4p^3(1-p)] + 15[p^2 + 2p(1-p)]^2$\\
& $+ 24[p^5 + 5p^4(1-p)] + 20[p^3 + 3p^2(1-p)][p^2 + 2p(1-p)]$\\

\hline\hline

$1 + \lambda_{4}^{\rm (core/halo)}$ & 
$1 + 6 f_c^2 [ (1 - p_c)^2 + 2 p_c (1-p_c) ] 
   + 8 f_c^3 [ (1 - p_c)^3 + 3 p_c (1-p_c)^2]$\\
& $+ 9 f_c^4 [ (1 - p_c)^4 + 4 p_c (1-p_c)^3]$\\

\hline

$1 + \lambda_{5}^{\rm (core/halo)}$ &
$1 + 10 f_c^2 [ (1 - p_c)^2 + 2 p_c (1-p_c) ]
   + 20 f_c^3 [ (1 - p_c)^3 + 3 p_c (1-p_c)^2]$\\
& $+ 45 f_c^4 [ (1 - p_c)^4 + 4 p_c (1-p_c)^3]
   + 44 f_c^5 [ (1 - p_c)^5 + 5 p_c (1-p_c)^4]$\\

\hline
\end{tabular}
\end{table}

By making use of the degrees of coherence $\lambda_{4}^{\rm (core/halo)}$ and $\lambda_{4}^{\rm (GL)}$ for four-pion BEC in the GL formula given in Table~\ref{tab12}, we can calculate four-pion BEC for the LHCb data as follows. First, we present the formula based on the core/halo model,
\begin{eqnarray}
\left.\frac{N^{(4\pi)}}{N^{\rm BG}}\right|_{\rm core/halo} = [1 + l_{42} e^{-RQ} + l_{43} e^{-\frac 32RQ} + l_{44} e^{-2RQ}]\times (C(1+\delta))^2,
\label{eq10}
\end{eqnarray}
where $l_{42}=6 f_c^2 [ (1 - p_c)^2 + 2 p_c (1-p_c) ]$, $l_{43}=8 f_c^3 [ (1 - p_c)^3 + 3 p_c (1-p_c)^2]$, and $l_{44}= 9 f_c^4 [ (1 - p_c)^4 + 4 p_c (1-p_c)^3]$. Second the formula based on the GL formula with geometrical two-component is given as follows:
\begin{eqnarray}
\left.\frac{N^{(4\pi)}}{N^{\rm BG}}\right|_{\rm GL} &\!\!=&\!\!
[1 + 6(p^2\cdot D + 2p(1-p)\cdot FD)
+ 8(p^3\cdot D^{1.5} + p\cdot D^{1/2}\cdot 3p(1-p)\cdot FD)\nonumber\\
&\!\! + &\!\! 6(p^4\cdot D^{2} + 2p^2\cdot D\cdot 2p(1-p)\cdot FD)
+ 3(p^2\cdot D + 2p(1-p)\cdot FD)^2]
\times \left({\rm LRC_{(Gauss)}}\right)^2,\nonumber\\
\label{eq11}
\end{eqnarray}
where $D=1/(1+(R_1Q)^2)^{2}$ and $FD=1/(1+(R_2Q)^2)^{1.5}$. Note that the power of $D$ is reflects the power of $p=A/\langle n\rangle$. The power of ${\rm LRC_{(Gauss)}}$ is related to $p^4$ as $4/2=2$. 

By using the estimated parameters for three-pion BEC in Tables~\ref{tab1} and \ref{tab4}, we can compute the four-pion BEC. The results are shown in Fig.~\ref{fig6}. ${\rm LRC_{(Gauss)}}$ is computed using the parameters for three-pion BEC in Table~\ref{tab4}. We expect this empirical analysis to be performed by the LHCb Collaboration in the near future. In that case, we will be able to examine the usefulness of quantum optics in hadron physics (see Ref.~\cite{UA1-MinimumBias:1991afz}).


\end{document}